\documentclass[
reprint,
secnumroman,
amssymb, amsmath,
aip,jcp,
bibtex
]{revtex4-1}

\usepackage[english]{babel}
\babeltags{en=english}
\babeltags{eng=english}

\usepackage{CJK}

\usepackage{graphicx}
\usepackage{dcolumn}
\usepackage{subfig}
\usepackage{siunitx}
\usepackage[usenames,dvipsnames,svgnames,table]{xcolor}

\expandafter\ifx\csname package@font\endcsname\relax\else
\expandafter\expandafter
\expandafter\usepackage
\expandafter\expandafter
\expandafter{\csname package@font\endcsname}%
\fi

\begin{document}
\begin{CJK*}{GB}{}
\title{Particle-based membrane model for mesoscopic simulation of cellular dynamics}
\author{Mohsen Sadeghi}
\thanks{Corresponding author}
\email{mohsen.sadeghi@fu-berlin.de}
\affiliation {Department of Mathematics and Computer Science, Freie Universit{\"a}t Berlin, Arnimallee 6, 14195 Berlin, Germany}
\author{Thomas R. Weikl}
\email{thomas.weikl@mpikg.mpg.de}
\affiliation{Department of Theory and Bio-Systems, Max Planck Institute of Colloids and Interfaces, Science Park Golm, 14424 Potsdam, Germany}
\author{Frank No\'e}
\thanks{Corresponding author}
\email{frank.noe@fu-berlin.de}
\affiliation {Department of Mathematics and Computer Science, Freie Universit{\"a}t Berlin, Arnimallee 6, 14195 Berlin, Germany}
\begin{abstract}
	We present a simple and computationally efficient coarse-grained and solvent-free model for simulating lipid bilayer membranes. In order to be used in concert with particle-based reaction-diffusion simulations, the model is purely based on interacting and reacting particles, each representing a coarse patch of a lipid monolayer. Particle interactions include nearest-neighbor bond-stretching and angle-bending, and are parameterized so as to reproduce the local membrane mechanics given by the Helfrich energy density over a range of relevant curvatures. In-plane fluidity is implemented with Monte Carlo bond-flipping moves. The physical accuracy of the model is verified by five tests: (i) Power spectrum analysis of equilibrium thermal undulations is used to verify that the particle-based representation correctly captures the dynamics predicted by the continuum model of fluid membranes. (ii) It is verified that the input bending stiffness, against which the potential parameters are optimized, is accurately recovered. (iii) Isothermal area compressibility modulus of the membrane is calculated and is shown to be tunable to reproduce available values for different lipid bilayers, independent of the bending rigidity. (iv) Simulation of two-dimensional shear flow under a gravity force is employed to measure the effective in-plane viscosity of the membrane model, and show the possibility of modeling membranes with specified viscosities. (v) Interaction of the bilayer membrane with a spherical nanoparticle is modeled as a test case for large membrane deformations and budding involved in cellular processes such as endocytosis. The results are shown to coincide well with the predicted behavior of continuum models, and the membrane model successfully mimics the expected budding behavior. We expect our model to be of high practical usability for ultra coarse-grained molecular dynamics or particle-based reaction-diffusion simulations of biological systems.
\end{abstract}
\maketitle
\end{CJK*}
\section{Introduction}
Lipid bilayer membranes are integral parts of the machinery of living cells. Apart from the obvious role of providing a mechanical and chemical barrier for the cell, they form the boundary of nearly all the organelles inside the eukaryotic cells and also take part in cellular functions such as signal transduction \cite{Alberts2015}. Biologically relevant processes at membranes, such as protein recruitment and insertion, assembly of protein scaffold at membranes, and membrane remodeling often involve spatial scales from tens to hundreds of nanometers, and time scales from milliseconds to minutes \cite{Alberts2015}. As an example, consider endocytosis and exocytosis at plasma membranes \cite{Haucke2011}. While all-atom molecular simulations are extremely successful for the study of individual macromolecules and small complexes\cite{LiRouxSchulten_NSMB14_VoltageGating, Lindorff-Larsen2011a, Lange2008, BuchFabritiis_PNAS11_Binding, Kohlhoff2014, PlattnerNoe_NatComm15_TrypsinPlasticity, WieczorekEtAl_NatComm16_MHCII} and can reach thermodynamics and kinetics at very long timescales with the aid of enhanced sampling methods and Markov state modeling \cite{Wu2014,Tiwari_PNAS14_KineticsProteinLigandUnbinding,Wu2016,TiwaryParrinello_PRL14_MetadynamicsDynamics,PlattnerEtAl_NatChem17_BarBar,Tiwary2017,Chodera2011,Doerr2016,Kohlhoff2014,Donati2017a}, they have severe limitations in terms of system sizes that can be sampled exhaustively \cite{Yu2016}. Even for the simple case of equilibrating micron-sized biomembranes, a blind scale up in all-atom molecular dynamics would be out of reach of computational power for decades to come \cite{Deserno2009}. To fill this computational gap, and gain insights into cellular processes, the development and application of coarse-grained models is an important aspect of computer simulation. A particularly promising framework to model cellular signaling processes at membranes, involving space exclusions and specific geometries found at membrane scaffolds, is particle-based reaction-diffusion (PBRD) simulation \cite{ZonTenWolde_PRL05_GFRD, ErbanChapman_PhysBiol09_StochasticReactionModelling,Klein2014,Hoffmann2014}, especially the so-called interacting-particle reaction-diffusion (iPRD) models that include interaction forces between particles \cite{SchoenebergNoe_PlosOne13_ReaDDy, BiedermannEtAl_BJ15_ReaddyMM, Vijaykumar2015, Gunkel2015,SchoenebergUllrichNoe_BMC14_RDReview}. The particles in such models typically represent entire proteins, protein domains or metabolites, and thus represent a spatial resolution of a few nanometers. Despite the success of such models in simulating cellular signal transduction processes \cite{SchoenebergEtAl_BJ14_PhototransductionKinetics, Ullrich2015, Gunkel2015, SchoenebergEtAl_NatComm17_SNX9}, these approaches are missing membrane mechanics in order to be able to model signaling at biomembranes. In spite of the extensive research on membrane models, there is arguably no readily usable model at the same scale, that is suited to be integrated into such a particle-based reaction-diffusion framework. Especially when it is required that the model be easily tunable, robust, and yet computationally efficient.

Bilayer membranes have been the subject of computer simulations for more than three decades \cite{Shillcock2006, Marrink2009, Noguchi2009}. Apart from all-atom simulations based on general purpose \cite{Marrink2001} or specifically developed force-fields \cite{Lindahl2000}, a vast variety of coarse-grained computational models developed for bilayer membranes exist (see \cite{Shillcock2006,Venturoli2006} and references therein for an overview). While we don't aim to provide a comprehensive review of all coarse-grained membrane models, it is useful to look at important modeling approaches and categorize them based on the level of coarse-graining achieved. This way, it becomes clear where the proposed model fits, and how it provides features suitable for its integration in iPRD simulations.

The first level of coarse-graining is achieved through grouping a specific set of atoms in lipid molecules into interaction centers and building effective force-fields \cite{Saunders2013}. The well-known MARTINI force-field falls into this category \cite{Marrink2004,Marrink2007,Arnarez2015}. Though considerably reducing the number of particles, this approach is still only suitable for simulating relatively small scale processes \cite{Marrink2013}. More recently, Srivastava and Voth devised a general approach for developing similar coarse-grained models for membranes composed of specific lipid molecules or lipid mixtures. Their approach consists of calibrating the interaction potentials to result in desired macroscopic mechanics and structural properties \cite{Srivastava2013}. Simunovic \textit{et al} employed this model to successfully simulate membrane remodeling by curvature-inducing proteins \cite{Simunovic2013a,Simunovic2013,Simunovic2015}. The next step in coarse-graining is to develop ``bead'' models, in which various lipid molecules are represented not by interaction groups pertaining to their atomistic representations, but by a small chain of generic particles \cite{Goetz1998,Farago2003,Cooke2005,Cooke2005a,Yamamoto2003,Wang2005,Lenz2005,Shillcock2002, Revalee2008, Huang2012}. A major challenge with these models is the choice of interaction potentials \cite{Cooke2005}. A higher level of coarse-graining pertains to one-particle-thick models in which the curvature elasticity is recovered through orientation-dependent pairwise interactions  \cite{Drouffe1991,Kohyama2009,Ballone2006,Yuan2010}. Drouffe \textit{et al.} pioneered this approach, and showed that through these orientation-dependent interactions, stable membranes and vesicles can form\cite{Drouffe1991}; though with the side-effect of predicting considerably low bending rigidities. To control the bending rigidity, Kohyama proposed a model in which local curvature of the membrane affects the particle-particle interactions \cite{Kohyama2009}. Ayton and Voth developed a systematic approach for parameterizing the interaction potential in their EM-DPD, and later, EM2 membrane models \cite{Ayton2002,Ayton2002c,Ayton2006,Ayton2007}. They performed detailed atomistic simulations, and employed energy equivalence in bending and bulk expansion/contraction modes to obtain optimal parameters for the mesoscopic model. They further applied these models in the study of membrane remodeling \cite{Ayton2007,Ayton2009}. From a different perspective, one-particle-thick models are also approached as discretized continuum models. Triangulated-surface models developed by Gompper and Kroll \cite{Gompper1997, Gompper1999, Gompper2000}, and Noguchi and Gompper \cite{Noguchi2005a, Noguchi2005} follow such an approach, and instead of relying on pairwise orientation-dependent potentials, uses angle-bending potentials between neighboring triangles to directly reproduce the curvature elasticity in a discretized model. Bahrami \textit{et al.} used a similar model to study interaction of nanoparticles with membranes \cite{Bahrami2012a,Bahrami2014} and formation of membrane tubules \cite{Bahrami2017}. Atilgan and Sun also incorporated the effect of transmembrane proteins into a triangulated model\cite{Atilgan2007}. As the dimensions and areas of triangular elements in these models can fluctuate, it is common practice with triangulated-surface models to utilize additional area-preserving constraints to control the surface area of the membrane. Another approach is to include the elasticity of an underlying continuous membrane into a particle-based description through potentials that depend on local surface fitting \cite{Noguchi2006,Noguchi2011}. Finally, the continuum description with curvature elasticity can also be solved numerically through available finite element methods developed for thin shell mechanics \cite{Feng2006}. In effect, these approaches substitute particles with computational nodes of a discretized continuum model.

In this paper, we introduce a novel coarse-grained membrane model which employs a two-particle-thick description of the bilayer membrane, with each particle effectively representing a patch of lipids on each leaflet. This is a minimal structure that allows for flexibility in modeling interactions of biomolecules with the membranes. The model relies on simple bond-stretching and angle-bending potentials in a dynamically updated bonded network, and thus, provides enhanced computational efficiency through the exclusion of non-bonded pairwise interactions. The proposed model is essentially an elastic membrane model, comparable to triangulated models, with the difference that the desired elastic properties are reproduced through simple bonded interactions in contrast to complicated orientation- or curvature-dependent potentials. Through a parameter-space optimization scheme, these interactions are easily tuned to reproduce membranes with desired elastic properties. The ultimate aim of developing such a model is to include it in large-scale simulations of cellular dynamics, and to specifically use it for studying cellular signal transduction using iPRD models. The computer experiments laid out in the following are designed to show that, despite its relative simplicity, inexpensive simulations done with the model very well reproduce expected behavior in terms of thermal undulations, area compressibility, in-plane viscosity, and budding under the influence of external forces.

\section{The Model}
\begin{figure}[t]
	\includegraphics[width=\columnwidth]{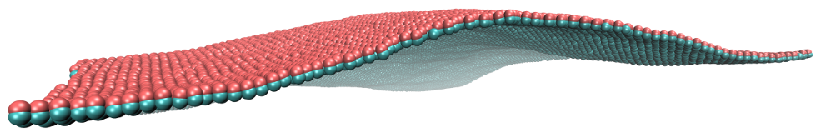}
	\textbf{(a)}
	\includegraphics[width=\columnwidth]{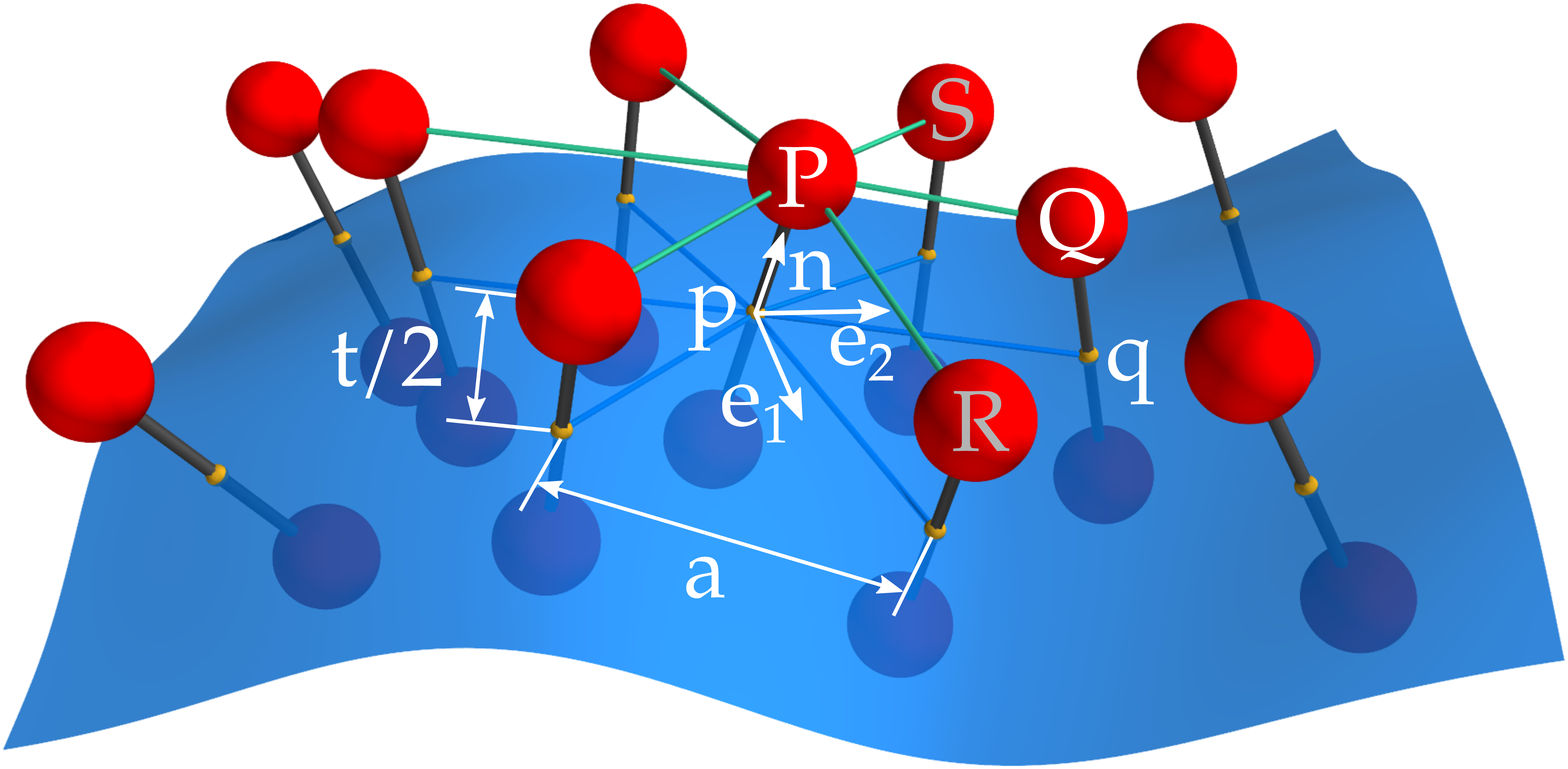}
	\textbf{(b)}
	\caption{(a) Snapshot of the proposed membrane model with particles forming top and bottom leaflets in red and cyan color, respectively. (b) Local surface geometry of the mid-surface in an arbitrary state of deformation (blue surface) with a collection of particle dimers whose positions are dictated by the mid-surface geometry. Distances and angles between these particles are used in order to probe the local curvature and relate between the particle model and continuum description of membrane mechanics.}
	\label{fig:model_1}
\end{figure}
As shown in Fig. \ref{fig:model_1}a, two close-packed lattices of particles correspondingly represent the two leaflets of the membrane in this model. The elastic energy density contributed to the membrane is usually expressed in terms of the local curvature of the mid-surface of the bilayer. We aim to avoid computing complex potential functions based on numerically obtained local curvature values. Thus, only bond-stretching and angle-bending interactions amongst nearest neighbor particles are considered. Considering an arbitrarily curved membrane, and based on its local surface geometry, relative configuration of particles, and the resulting bond lengths and angles are obtained. An effective energy density pertaining to bonded interactions is thus calculated, and compared with the curvature elasticity modeled via the Helfrich energy density to parameterize the interaction potentials.
\subsection{Curvature elasticity of the bilayer membrane}  
The Helfrich energy density of a curved fluid bilayer membrane is expressed as \cite{Canham1970,Helfrich1973,Evans1974}
\begin{equation}
\label{eq:Helfrich}
f_{\mathrm{H}} = 2 \kappa (H - H_0) ^ 2 + \bar{\kappa} G
\end{equation}
in which the constant $\kappa$ is the bending rigidity or splay modulus of the membrane and $\bar{\kappa}$ is its Gaussian curvature rigidity or saddle-splay modulus. $H$ and $G$ represent the mean and Gaussian curvatures, respectively, which are defined based on the principal curvatures, $c_1$ and $c_2$, as $H=\left(c_1 + c_2\right)/2$ and $G=c_1 c_2$. $H_0$ is the spontaneous mean curvature of the membrane, corresponding to a local curvature that is induced in the membrane not by external forces, but by internal effects such as the geometry of phospholipid molecules \cite{Rozycki2015}.
\subsection{Differential geometry of the particle-based membrane model}
\label{sec:diff_geom}
In this model, in which the membrane is effectively composed of ``particle dimers'', i.e. pairs of particles belonging to the top and bottom leaflets, a hypothetical mid-surface is assumed to lie halfway between the particle dimers. Inspired by classical continuum shell theories, we assume that bending of the double layer deforms it such that a normal vector originating from a point $\mathrm{p}$ on the mid-surface, pointing to a particle $\mathrm{P}$ on the upper or lower layer, remains perpendicular to the mid-surface, independent of the state of deformation (see Fig. \ref{fig:model_1}b). Thus, the position of the particle $\mathrm{P}$ is always given as $\mathbf{r}_\mathrm{P} = \mathbf{r}_\mathrm{p} \pm \frac{t}{2}\mathbf{n}$, where $\mathbf{n}$ is the normal vector of the mid-surface at point $\mathrm{p}$, $t$ is the thickness of the membrane, and the plus and minus signs correspond to particles on the top or bottom leaflets, respectively. Without loss of generality, we focus on particles positioned on the top leaflet for the following derivations. For two neighboring particles $\mathrm{P}$ and $\mathrm{Q}$, corresponding mid-surface projections are considered to be $\mathrm{p}$ and $\mathrm{q}$, given by local coordinates, $\mathbf{r}_\mathrm{p} = \left(0, 0\right)$ and $\mathbf{r}_\mathrm{q} = \left(u^1, u^2\right)$, respectively (Fig. \ref{fig:model_1}b). Thus, the set of coordinates, $u^1$ and $u^2$, provide a local parameterization of the mid-surface in the vicinity of point $\mathrm{p}$. For point $\mathrm{q}$, this description can be approximated through a second-order Taylor expansion:
\begin{equation}
\label{eqn:Taylor1}
\begin{split}
\mathbf{r}_\mathrm{q} &\approx \mathbf{r}_\mathrm{p} + u^\mu\mathbf{e}_\mu + \frac{1}{2}u^\mu u^\nu\mathbf{e}_{\mu,\nu}\\&=\mathbf{r}_\mathrm{p} + u^\mu\mathbf{e}_\mu + \frac{1}{2}u^\mu u^\nu\left(\Gamma^\sigma_{\mu\nu}\mathbf{e}_\sigma+b_{\mu\nu}\mathbf{n}\right)
\end{split}
\end{equation}
where $\mathbf{e}_\mu = \partial_\mu{\mathbf{r}}$ are the base vectors for the tangent space at point $\mathrm{p}$, $\Gamma^\sigma_{\mu\nu}=\mathbf{e}_{\mu,\nu}\cdot\mathbf{e}^{\sigma}$ are the Christoffel symbols of the second kind and $b_{\mu\nu}=\mathbf{e}_{\mu,\nu}\cdot\mathbf{n}$ are the components of the second fundamental form tensor \cite{Frankel2012}. It is to be noted that summation convention between a pair of upper and lower indices is used here. Similarly, another Taylor expansion can be used to approximate the normal vector at point $\mathrm{q}$, making it possible to express the position of particle $\mathrm{Q}$ with respect to particle $\mathrm{P}$ as:
\begin{equation}
\label{eqn:PQ1}
\begin{split}
\mathbf{r}_{\mathrm{PQ}} &= \mathbf{r}_\mathrm{Q} - \mathbf{r}_\mathrm{P} \\&\approx \left[u^\sigma + \frac{1}{2}\Gamma^\sigma_{\mu\nu}u^\mu u^\nu-\frac{h}{2}\,b_{\mu\nu}g^{\nu\sigma}u^\mu \right]\mathbf{e}_\sigma\\&+\frac{1}{2}b_{\mu\nu}u^\mu u^\nu \mathbf{n}
\end{split}
\end{equation}
in which $g_{\mu\nu}=\mathbf{e}_\mu\cdot\mathbf{e}_\nu$ is the metric tensor and we have $g^{\mu\sigma}g_{\sigma\nu}=g_{\mu\sigma}g^{\sigma\nu}=\delta^\mu_\nu$ with $\delta^\mu_\nu$ being the Kronecker's delta. For the purpose of calculating partial derivatives of the normal vector, the Weingarten's formula, $\mathbf{n}_{,\mu} = -b^\nu_\mu\mathbf{e}_\nu = -b_{\mu\nu}g^{\nu\sigma}\mathbf{e}_\sigma$ has been used \cite{Frankel2012}. It is noteworthy to mention that first order partial derivatives of the normal vector contain second order derivatives of the position vector, $\mathbf{r}$, through the inclusion of the $b_{\mu\nu}$ tensor components, effectively making the two approximations of the same order. The length of the vector $\mathbf{r}_{\mathrm{PQ}}$ as well as the angle it makes with the normal vector at point $\mathrm{p}$, are obtained by forming the following inner products,
\begin{equation}
\label{eqn:rPQ2}
\begin{split}
\left|\mathbf{r}_{\mathrm{PQ}}\right|^2 &= \mathbf{r}_{\mathrm{PQ}}\cdot\mathbf{r}_{\mathrm{PQ}}\\
&\approx \mathrm{I}(\mathbf{u})\left(1-\frac{t^2}{4}G \right) - \mathrm{I\!I}(\mathbf{u})\, t \left(1- \frac{t}{2}H\right)\\
&+ \frac{1}{4}\mathrm{I\!I}^2(\mathbf{u}) + C_1 - \frac{t}{2}C_2  + \frac{1}{4}C_3 
\end{split}
\end{equation}
and
\begin{equation}
\label{eqn:PQdotn}
\left|\mathbf{r}_{\mathrm{PQ}}\right|\left|\mathbf{n}\right|\cos \theta_{\mathrm{p P Q}} = -\mathbf{r}_{\mathrm{PQ}}\cdot\mathbf{n} \approx -\frac{1}{2}\mathrm{I\!I}(\mathbf{u})
\end{equation}
where,
\begin{equation}
\label{eqn:fund_form}
\begin{split}
\mathrm{I}(\mathbf{u}) &= g_{\mu\nu}u^\mu u^\nu\\
\mathrm{I\!I}(\mathbf{u}) &= b_{\mu\nu}u^\mu u^\nu
\end{split}
\end{equation}
are the first and second fundamental forms. The remaining parameters in eq. (\ref{eqn:rPQ2}) are defined as follows:
\begin{equation}
\label{eqn:PQ_params}
\begin{split}
C_1 &= \Gamma_{\mu\nu\sigma}u^\mu u^\nu u^\sigma\\
C_2 &= \Gamma^{\sigma}_{\mu\nu}b_{\sigma\gamma}u^\mu u^\nu u^\gamma\\
C_3 &= \Gamma^{\sigma}_{\mu\nu}\Gamma_{\mu^\prime\nu^\prime\sigma}u^\mu u^\nu u^{\mu^\prime} u^{\nu^\prime}
\end{split}
\end{equation}
where $\Gamma_{\mu\nu\sigma}=\Gamma^{\gamma}_{\mu\nu}g_{\gamma\sigma}$ are the Christoffel symbols of the first kind \cite{Frankel2012}. Up to this point, the derived equations hold in all local coordinate systems at point $\mathrm{p}$. A smart choice of the coordinate system can simplify the equations considerably. A perfect candidate is the locally tangent coordinate system, with the following implicit definition:
\begin{equation}
\label{eqn:tancoord}
u^\sigma = u^{\star \sigma} - \frac{1}{2}\Gamma^{\sigma}_{\mu\nu} u^{\star \mu}u^{\star \nu}
\end{equation}
in which $\left(u^{\star 1}, u^{\star 2}\right)$ are the new coordinates with the same origin at point $\mathrm{p}$, and the Christoffel symbols are calculated in the old coordinate system at point $\mathrm{p}$. It can be shown that in this coordinate system, Christoffel symbols vanish identically, and yet, because $ \left(\partial u^{\mu}/\partial u^{\star \nu}\right)_\mathrm{p} = \delta ^{\mu}_{\nu}$, first order length differentials as well as the first and second fundamental forms remain unchanged. 
\subsection{Parameter-space optimization of interaction potentials}
\label{sec:parameter_space_opt}
Now that we have obtained equations describing the relative configuration of model particles in an arbitrarily curved membrane (Eqs. \ref{eqn:rPQ2} and \ref{eqn:PQdotn}), we can select interaction potentials which are functions of $\left|\mathbf{r}_{\mathrm{PQ}}\right|$ and $\theta_{\mathrm{p P Q}}$, and calculate effective energy densities corresponding to arbitrary curvature states. In effect, we seek to obtain numerical values of the energy density arising from a specific set of interaction potentials, as a function of mid-surface curvature, prior to running an actual simulation with these potentials. As a simple choice, we assume that nearest neighbor particles on both top and bottom leaflets are connected via lateral bonds. Also, an angle-bending potential is assumed to exist for out-of-plane rotations of such bonds. These two bonded interactions are handled respectively with the following Morse-type bond-stretching and harmonic angle-bending potentials:
\begin{equation}
\label{eq:pot_model}
\begin{split}
U_{\mathrm{stretch}}\left( \left|\mathbf{r}_{\mathrm{PQ}}\right| \right) &= D_{\mathrm{e}}\left[1 - e^{-\alpha \left(\left|\mathbf{r}_{\mathrm{PQ}}\right| - a \right)}\right]^2\\
U_{\mathrm{bend}}\left(\theta_{\mathrm{P^{\prime} P Q} }\right) &= K_{\mathrm{b}}\left(\theta_{\mathrm{P^{\prime}PQ}} - \frac{\pi}{2}\right)^2
\end{split}
\end{equation}
where $a$ denotes the lattice parameter (or equilibrium separation of particles on each leaflet) and $\mathrm{P^{\prime}}$ is the particle residing on the bottom leaflet, forming a dimer with particle $\mathrm{P}$. The equilibrium angle is chosen to be $\pi/2$, which corresponds to angle-bending with respect to a flat membrane.

It is to be noted that this choice of bonded interactions, and the potentials to handle them, is by no means unique. The general procedure laid out here can be applied to many other choices, with the condition that geometric information can be extracted uniquely from the curvature of the mid-surface.

In order to calculate the effective energy density, an area element on the mid-surface corresponding to a set of interactions has to be defined. We propose Voronoi tessellation be used to do so in a systematic way. In the simple case of a hexagonal close-packed lattice of particles, Voronoi tessellation simply yields hexagons centered at particles' projections on the mid-surface. Though in general, the shape and area of elements corresponding to particle projections is a function of their coordination number. Especially considering the fact that the coordination number changes due to bond-flipping Monte Carlo moves that will be discussed in Sec. \ref{sec: bondflipping}. With such a definition for area elements, half of each lateral bond emanating from a particle $\mathrm{P}$, plus all the out-of-plane angles having it as the vertex, are included in one area element around particle $\mathrm{P}$. But without performing the simulation, we don't have a priori knowledge of the in-plane angle, $\psi$, that this star-shaped construct around each particle makes with the principal directions of the curvature of the mid-surface. Thus, in general, the calculated effective energy density depends on this in-plane angle. To compensate for this ambiguity, and avoid directional bias, the effective energy density is numerically averaged out over possible values of $\psi$. This way, the effective potential energy density is defined as:
\begin{equation}
\label{eqn:f_eq}
f_{\mathrm{eff}} =\left \langle\frac{\frac{1}{2}\sum {U_{\mathrm{stretch}}} + \sum {U_{\mathrm{bending}}}}{\Delta A}\right\rangle_{\psi}
\end{equation}
where the summations are carried out for all interactions corresponding to one particle and $\Delta A$ denotes the area element. The same procedure applies to the pair particle, $\mathrm{P}^{\prime}$, which lies on the bottom leaflet, and the corresponding energy density is simply added to $f_{\mathrm{eff}}$.

The chosen interaction potentials given in eq. \ref{eq:pot_model} contain a set of parameters, $D_{\mathrm{e}}$, $\alpha$, and $K_{\mathrm{b}}$. In order to obtain optimal values for these parameters, a dimensionless error measure is defined as
\begin{equation}
\label{eqn:err_f_eq}
e = \frac{\int dc_1 \int dc_2 \,\,\left(f_{\mathrm{eff}}-f_{\mathrm{H}}\right)^2}{\int dc_1 \int dc_2 \,\,f_{\mathrm{H}}^2}
\end{equation}
in which the integration is carried out in the mid-surface curvature space spanned by its principal curvatures, $c_1$ and $c_2$. The integration range is arbitrary, and corresponds to the range of curvatures that have practical relevance. Minimizing this error measure with respect to potential parameters yields their optimal values.
\subsection{Additional interactions}
The bond-stretching and angle-bending interactions described in Sec. \ref{sec:parameter_space_opt} only serve to reproduce the desired curvature elasticity of the membrane. In addition to these interactions, other potentials can be added to the model for different geometrical or mechanical considerations, as long as they do not perturb the effective energy density described by eq. \ref{eqn:f_eq}.

Most importantly, a harmonic potential is added between particles in each dimer. This potential is described by the expression $U_\mathrm{thickness}\left(t\right)=K_{\mathrm{t}} \left(t - t_0\right)^ 2$ where $t$ is the center-to-center distance of the particles, and $t_0$ is the prescribed equilibrium thickness of the membrane. Addition of this potential is necessary to hold the two leaflets together. The potential strength, $K_{\mathrm{t}}$, should be chosen high enough to preserve the thickness variations within a reasonable range, and to prevent the thermal motion of the particles to cause them to flip between the leaflets. Note that this potential may also be calibrated with respect to the actual stiffness of bilayer membranes across their thickness.

Also, with the set of interactions described so far, volume exclusion is only present between neighboring membrane particles, and in principle, non-neighboring particles can interpenetrate. In simulations where this issue arises, a non-bonded interaction, such as harmonic repulsion, can additionally be included.

\subsection{Bond-flipping moves}\label{sec: bondflipping}
The membrane model developed so far is based on a fixed topology of bonded interactions, and thus, pertains to a two-dimensional solid. In contrast, lipid bilayer membranes are two-dimensional fluids in which lipid molecules can freely diffuse laterally, and this fluidity is essential for membrane remodeling \cite{Seifert1997}. Following a scheme commonly used in triangulated membrane models \cite{Noguchi2005,Noguchi2005a}, the in-plane fluidity is introduced to the model via bond-flipping Monte Carlo moves. In a quadrilateral formed by four neighboring particles (e.g. $\mathrm{PRQS}$ in Fig. \ref{fig:model_1}b), swapping of one diagonal bond ($\mathrm{PQ}$) with the other ($\mathrm{RS}$) is proposed with a frequency $\phi$ during the simulation. This proposed move is accepted with the Metropolis-Hastings probability of $\exp \left[-\beta \left(E_{\mathrm{new}} - E_{\mathrm{old}}\right) \right]$ where $E_{\mathrm{old}}$ and $E_{\mathrm{new}}$ are the corresponding potential energies of the system in the old and new topologies, and $\beta = 1/kT$ with $k$ being the Boltzmann constant and $T$ the temperature. Whenever a proposed bond-flipping move is accepted, the topology of the model is updated to remove the currently existing $\mathrm{PQ}$ bond (and its accompanying angles), and substitute them with a newly formed $\mathrm{RS}$ bond (and corresponding angles). Introduction of these Monte Carlo moves, which favor flipping bonds under tension to lower energy ones, results in a net energy loss. In a simulation in the canonical ensemble, this lost energy will be compensated by the thermostat, which is the same as extracting work and adding equal amount of heat to the system. Thus, in effect, this is an entropy production mechanism comparable to viscous loss.

The frequency, $\phi$, with which the bond-flipping moves are proposed, acts as a control parameter for the model. The in-plane dynamics of model particles, which determines kinetic properties such as the effective in-plane diffusion and surface viscosity of the membrane, can be manipulated via the frequency of bond-flipping moves. We expect the membrane to assume more fluidity and faster in-plane dynamics with increasing values of $\phi$. This assumption will be put to quantitative test in Sec. \ref{sec:shear_flow}.
\section{Simulation details}
\subsection{Parametrization}
\begin{figure}[t]
	\includegraphics[width=\columnwidth]{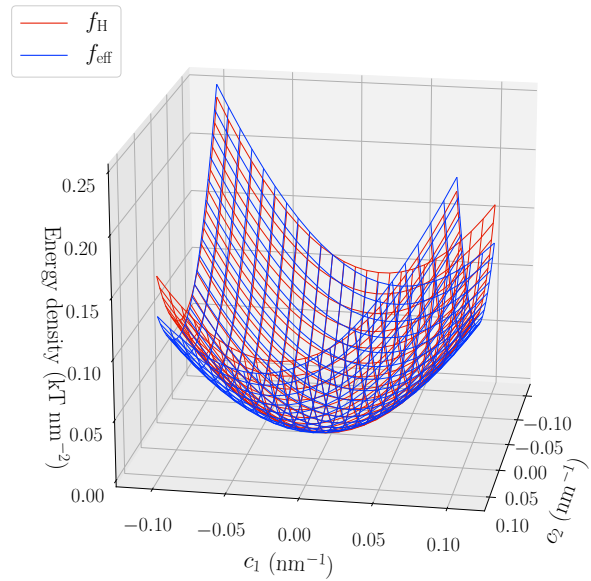}
	\caption{Comparison between the Helfrich ($f_{\mathrm{H}}$) and the effective energy density of the proposed model ($f_{\mathrm{eff}}$) with the optimal parameters for a hexagonal membrane model with the lattice parameter of \SI{10}{\nano\meter}.}
	\label{fig:eq_pot_1}
\end{figure}
In order to implement the parameter-space optimization procedure explained in Sec. \ref{sec:parameter_space_opt}, it is necessary to have elastic constants of a bilayer membrane, namely $\kappa$ and $\bar{\kappa}$, as input. Experimental determination of the bending rigidity, $\kappa$, is based on the two general approaches of monitoring fluctuations or by pulling out tethers and measuring the forces involved \cite{Nagle2013,Dimova2014}. As expected, the value of the bending rigidity depends on temperature, as well as the composition of the bilayer membrane. The range \SI{10}{} $kT$ to \SI{40}{} $kT$ is usually quoted \cite{Nagle2013,Dimova2014}. Since we are presently not focusing on simulating a membrane with a specific composition, the value of \SI{20}{} $kT$ is chosen for the bending rigidity \cite{Marsh2006}. On the other hand, the saddle-splay or Gaussian curvature modulus has been elusive to experimental determination. This is a direct consequence of the Gauss-Bonnet theorem which limits changes in the integral of the saddle-splay energy term to cases where the topology of the membrane changes. Methods based on membrane buckling have been proposed to calculate Gaussian curvature modulus in simulations \cite{Hu2012,Hu2013}, but they are not experimentally applicable. In order to obtain a realistic value for our simulations, the typical ratio $\bar{\kappa}/\kappa = -0.8$ is used to obtain the Gaussian curvature modulus of \SI{-16}{} $kT$ \cite{Marsh2006,Hu2012}.

For a given set of potential parameters, namely $D_{\mathrm{e}}$, $\alpha$, and $K_{\mathrm{b}}$ in eq. \ref{eq:pot_model}, the error measure in eq. \ref{eqn:err_f_eq} is obtained through numerical integration. The integration range for both principal curvatures is chosen to be \SIrange{-0.1}{0.1}{\per\nano\meter}. In order to calculate the integrand at each integration point, the expression given in eq. \ref{eqn:f_eq} for the $f_{\mathrm{eff}}$ is used. The averaging over the angle $\psi$ is also carried out numerically with fine \ang{1} divisions. Using the $C_z$ symmetry of the star-shaped bonded construct around each particle, where $z$ is the coordination number, this averaging needs only to be done over one $2\pi/z$ interval. We have used the quasi-Newton BFGS algorithm \cite{Knoll2004} from the SciPy optimize package for numerical minimization of the numerically calculated error measure. The iterative minimization algorithm is repeated several times, with randomly picked initial guesses in the parameter space, to help the algorithm converge to a global minimum.

The minimization process is performed in a bounded domain, with prescribed ranges for potential parameters. In general, for a given set of membrane elastic constants, and by manipulating potential parameter bounds, different ``families'' of optimal parameters are obtained. While these parameter families should theoretically produce the same curvature elasticity, the following considerations can be applied to favor some over the rest:
\begin{itemize}
	\item {Imposition of other physical properties on the model: Having different parameter families provides the flexibility to calibrate the model based on other physical properties, while preserving the curvature elasticity. The area compressibility modulus will be addressed as an example in Sec. \ref{sec:area_comp}.}
	\item {Technical necessities of the simulation: For example, a model limited to bond-stretching interactions with $D_{\mathrm{e}} > 0$ and $K_{\mathrm{b}} = 0$ is a valid result of the parameter-space optimization. But a double-layer model built thus would not resist the rotation of particle-dimers, resulting the two leaflets to interpenetrate.}
\end{itemize}
\begin{table}[h]
	\centering
	\caption{
			Optimized potential parameters for models with different lattice parameters and the same coordination number of 6. For the \SI{10}{\nano\meter} model, three families of parameters (designated by a, b, and c superscripts) are given.}
	\begin{tabular}{*4{>{\centering\arraybackslash}p{0.2\columnwidth}}}
		\toprule
		$a \, (\mathrm{nm})$ & $D_{\mathrm{e}} \, (\mathrm{kJ} \, \mathrm{mol} ^ {-1})$ & $\alpha \, (\mathrm{nm}^{-1})$ & $K_{\mathrm{b}} \, (\mathrm{kJ} \, \mathrm{mol} ^ {-1})$\\
		\hline
		10.0$^{(a)}$ & 17.98 & 0.120 & 17.34\\
		10.0$^{(b)}$ & 64.59 & 0.120 & 12.39\\
		10.0$^{(c)}$ & 111.2 & 0.120 & 7.433\\
		15.0 & 23.82 & 0.066 & 20.19\\
		20.0 & 24.41 & 0.050 & 22.92\\
		\hline
	\end{tabular}
	\label{tab:table_param}
\end{table}

Tab. \ref{tab:table_param} shows the results of parameter space optimizations performed for membrane models with different lattice parameters and the same coordination number of 6. To demonstrate the flexibility to choose different parameter families, for the \SI{10}{\nano\meter} model, three sets of parameters are given. These three families, that balance the energy differently between bond-stretching and angle-bending interactions, are produced by putting different bounds on the angle-bending stiffness, $K_{\mathrm{b}}$. For each case, the resultant potential parameters can be used to calculate the effective energy density as a function of mid-surface curvature. Fig. \ref{fig:eq_pot_1} shows a comparison between the Helfrich energy density and the optimized effective energy density in the curvature space, for the case of a model with the lattice parameter of \SI{10}{\nano\meter}, using the parameters given in the first row of Tab. \ref{tab:table_param}. Indeed, the presently parameterized potential agrees well with the Helfrich energy density over a relatively wide curvature range. 

It is to be noted that the elastic constants chosen here, and the resulting potential parameters given in Tab. \ref{tab:table_param}, serve as an example to illustrate the applicability of the model, where the general procedure of calculating the effective energy density and parameter-space optimization can be applied with any choice of $\kappa$ and $\bar{\kappa}$ values. This offers the flexibility of modeling different membranes via the same parameterization process.

In addition, for the harmonic potential $U_{\mathrm{thickness}}$, the strength of $K_{\mathrm{t}}=$ \SI{7.7}{\kilo\joule\per\mole\per\nano\meter\squared}, and equilibrium distance of $t=$ \SI{4.0}{\nano\meter} are respectively chosen.

Finally, the masses of the representative particles in the model are determined based on the effective surface density of bilayer membranes in equilibrium. For the simulations presented here, the case of a DPPC bilayer membrane is chosen, for which area per lipid is determined through atomistic simulations \cite{Edholm2005} to be \SI{0.640}{\nano\meter\squared}. This value results in a surface density of \SI{380.9}{\nano\gram\per\square\centi\meter} and individual particle mass of \SI{0.165}{\atto\gram}. A quick calculation shows that each of the representative particles in the model with the lattice parameter of \SI{10}{\nano\meter} thus accounts for about 140 DPPC molecules.
\subsection{Time integration}
In order to simulate tensionless membranes in thermal equilibrium, an extended systems dynamics  approach is used to derive equations of motion and devise the proper numerical integration scheme. The systematic approach developed by Martyna \textit{et al.} (the so-called MTK integrator) based on sequential application of discretized Liouville operators proved to be a robust way to achieve proper thermostatting and barostatting \cite{Martyna1994,Martyna1996,Tuckerman2006}. Thermostatting is achieved through Nos\'e-Hoover chains, and isotropic cell fluctuations are used for barostatting to achieve zero in-plane tension. With the mass chosen for individual particles and based on the optimized values calculated for potential parameters, the time step for the explicit integrator is chosen to be \SI{20}{\pico\second}.

It is to be noted that while the model developed so far has a well-defined physical length scale pertaining to the thickness of the membrane, attributing a singular time scale to it is not as straightforward. It is reasonable to assume that the motion of membrane particles is governed by two decoupled dynamics, respectively in the in-plane and out-of-plane directions. As was discussed in Sec. \ref{sec: bondflipping}, and as the results given in Sec. \ref{sec:shear_flow} will show, the in-plane dynamics of the model can be manipulated via changing the frequency of bond-flipping moves, and can thus be calibrated by comparing a resultant kinetic property, such as surface viscosity, with its respective experimental value. On the other hand, in the absence of any solvent effects, and with the deterministic MTK integrator used here, the out-of-plane dynamics is solely determined by the particle masses and the stiffness of the forcefield developed based on the scheme introduced in Sec. \ref{sec:parameter_space_opt}. As the forcefield is the outcome of the parameter-space optimization aiming to reproduce the desired membrane elasticity, the only remaining parameter is the mass of model particles. Choosing the deterministic MTK integrator has the advantage of putting the robustness of the model to the test where no prescribed damping is present, but has the side-effect of producing very fast out-of-plane dynamics with the current choice of surface density (see Sec. \ref{sec:thermal_undulations}). While the value of membrane surface density and the resulting particle mass can in principle be manipulated to control the time scale, doing so does not correspond to a meaningful physical setup. Achieving physically relevant out-of-plane dynamics pertaining to membrane patches suspended in a solvent is only possible through either implementing a suitable stochastic integrator, or including solvent effects explicitly or implicitly. This will be addressed in future applications of the model.
\subsection{Simulation code and visualization}
Mainly due to the fact that the implementation of bond-flipping Monte Carlo moves in available molecular dynamics software packages proved impractical, an in-house C++ code has been developed to handle the simulations. Visualization is done via the Visual Molecular Dynamics (VMD) software package \cite{HumphreyDalkeSchulten_JMG96_VMD}.
\section{Results and discussion}
\subsection{Thermal undulations}
\label{sec:thermal_undulations}
A lipid bilayer patch in thermal equilibrium undergoes significant out-of-plane thermal undulations \cite{Seifert1997}. These undulations can be studied from an statistical mechanics point of view to obtain energy distribution among different vibration modes. Considering a square membrane patch of side length $L$, its mid-surface can be parameterized as $\mathbf{r} = \left(x, y, h\left(x,y\right)\right)$, where $h$ is the height function (the so-called Monge description). Assuming that this membrane patch has periodic boundary conditions in the $x$ and $y$ directions, the height function can be expressed as a discrete Fourier series:
\begin{equation}
h = \sum_{m, n} \tilde{h}\left(\mathbf{q}_{m,n}\right)\exp\left(i\, \mathbf{q}_{m,n}\cdot\mathbf{r}\right)
\end{equation}
in which $\mathbf{q}_{m,n} = \frac{2\pi}{L}\, (m, n)$ is the wave vector. It can be shown that based on the Helfrich expression (Eq. \ref{eq:Helfrich}) the energy corresponding to each vibration mode is given by $\frac{1}{2}{\kappa L^2} q^4\, \tilde{h}(\mathbf{q})\,\tilde{h}^*(\mathbf{q})$, and thus, application of the equipartition theorem yields the power spectrum of thermal undulations as:
\begin{equation}
\label{eqn:thermal_und}
\frac{1}{L^2}\left\langle \tilde{h}(\mathbf{q})\,\tilde{h}^*(\mathbf{q}) \right\rangle = \frac{k T}{\kappa\,\left(qL\right)^4}
\end{equation}
The $q^{-4}$ power law is used as a test to observe if the particle-based model reproduces the continuum behavior dictated by the Helfrich energy density correctly. Also, fitting eq. \ref{eqn:thermal_und} to the results yields the value of the bending rigidity, $\kappa$.
\begin{figure}[t]
	\includegraphics[width=\columnwidth]{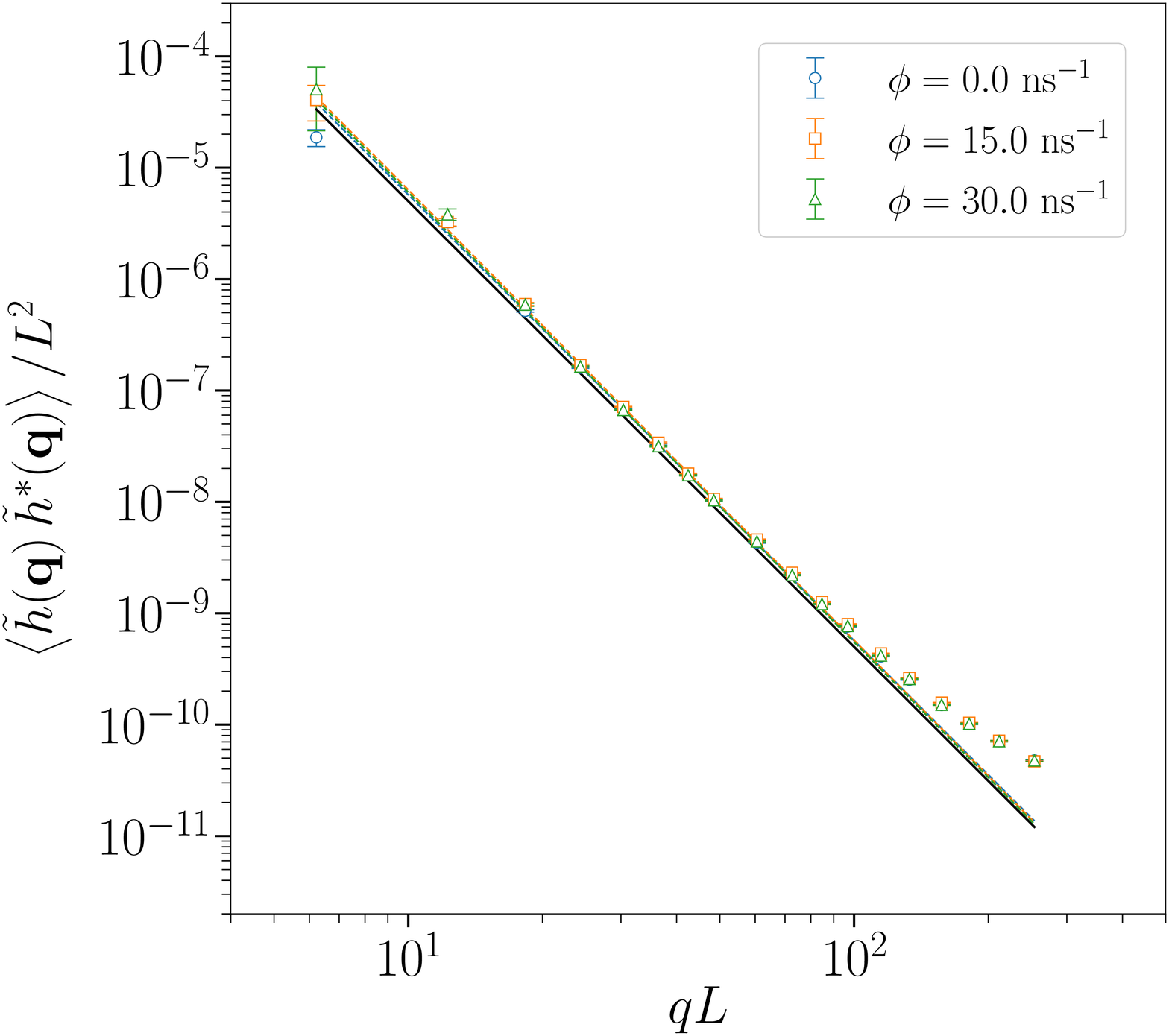}
	\caption{Power spectrum of thermal undulations of membrane patches with the lattice parameter of \SI{10}{\nano\meter} and different bond-flipping frequencies. All patches have the same lateral size of $\sim$\SI{1}{\micro\metre} and are equilibrated at \SI{298}{\kelvin}. Dashed lines are fits of the function $C \left(q L\right)^{n}$ to the data, whereas the solid black line is the prediction of the continuum model with the bending rigidity of \SI{20}{} $kT$, the same value used as an input for parameterizing the interaction potentials.}
	\label{fig:th_1}
\end{figure}
\begin{figure}[t]
	\includegraphics[width=\columnwidth]{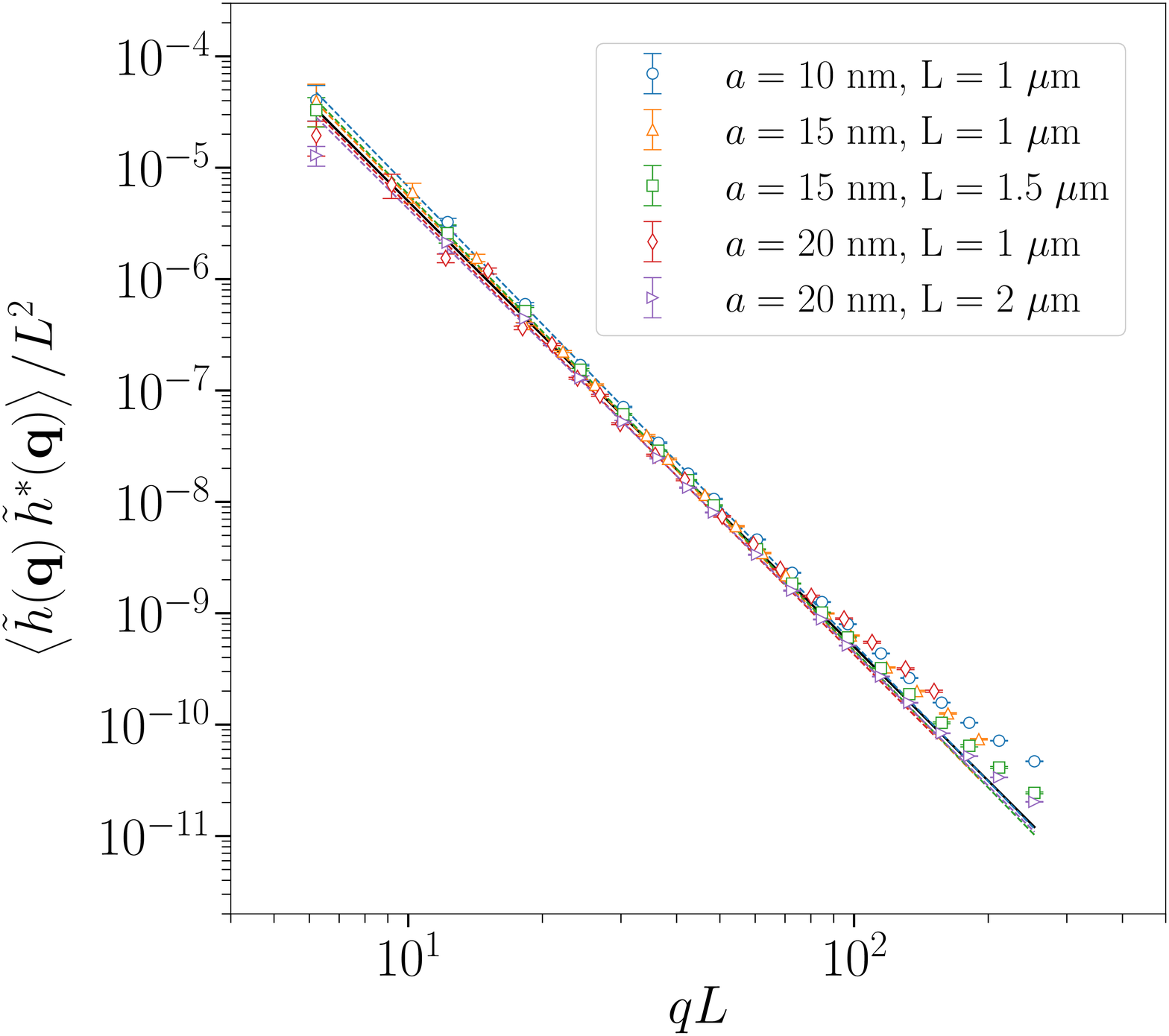}
	\caption{Similar to Fig. \ref{fig:th_1} for thermal undulations of membrane patches with various lattice parameters and lateral dimensions. All patches are simulated with the same bond-flipping frequency of \SI{15}{\per\nano\second}}.
	\label{fig:th_2}
\end{figure}

As a first experiment, membrane patches of approximately \SI{1}{\micro\meter} in size with the lattice parameter of \SI{10}{\nano\meter} are simulated at constant temperature of \SI{298}{\kelvin}. Potential parameters given in the first row of Tab. \ref{tab:table_param} are used. To ensure that the membrane patches have indeed been equilibrated, an estimate of the relaxation time of the system is required. Following Farago \cite{Farago2003}, two methods are used for gaining this estimate:
\begin{itemize}
 \item{Measuring the time it takes for the potential energy of the membrane to settle to fluctuations about an equilibrium value,}
 \item{Measuring the relaxation time of the longest wavelength in thermal undulations.}
\end{itemize}
Both measures give values in the \SI{1}{\micro\second} range, which signifies a rather fast out-of-plane dynamics. Thus, a total time of \SI{20}{\micro\second} is used for each simulation, out of which the second half is used for sampling observables. The resulting equilibrium trajectories are used to calculate a discrete height function defined on a constant spatial grid, and fast Fourier transform is used to extract its average power spectrum. This process is repeated for models with different bond-flipping frequencies, which are expected to have different in-plane fluidities. The results are depicted in Fig. \ref{fig:th_1}. The solid black line shows the prediction of the continuum model for the bending rigidity of \SI{20}{} $kT$. Our model reproduces the continuum behavior quite accurately, and also the bending rigidity is recovered very well. To further verify this, two sets of equations, $C \left(q L\right)^{n}$, and $\left (1/\kappa\right) \left(q L\right)^{-4}$ are fitted to the first 8 values of ${\left\langle \tilde{h}(\mathbf{q})\,\tilde{h}^*(\mathbf{q}) \right\rangle}/{L^2}$. It is to be noted that the expected continuum behavior only applies to high wavelength undulations and separation from the $q^{-4}$ behavior is to be expected at short wavelengths. The parameters for these fits are given in Tab. \ref{tab:table_1}. It is evident that for all cases, the $n=-4$ behavior of a continuum model is very well reproduced. Also, from the second fit, the magnitude of the effective bending rigidity of the membrane, $\kappa$, is obtained, and can be compared with the input value of \SI{20}{} $kT$ with good accuracy.

As the second test, the power spectrum of thermal undulations for models with different lattice parameters of \SIlist[list-units = single]{10;15;20}{\nano\meter} and the same bond-flipping frequency of $\phi=$ \SI{15}{\per\nano\second} are studied. To do a proper comparison, for lattice parameters other than \SI{10}{\nano\meter}, two cases are simulated. First, a square membrane patch with the same lateral dimension of \SI{1}{\micro\meter} is spanned with fewer particles at larger separations, and second, the same number of particles as the \SI{10}{\nano\meter} model are used, which form larger patches. The results are depicted in Fig. \ref{fig:th_2}. It is observed that increasing the lattice parameter in general has little effect on the ability of the model to reproduce continuum behavior. Tab. \ref{tab:table_2} gives the results of similar $C \left(q L\right)^{n}$ and $\left (1/\kappa\right) \left(q L\right)^{-4}$ fits to these data. Again, the expected $n=-4$ and $\kappa$ = \SI{20}{} $kT$ behavior is very well reproduced by the model. Comparing the results for membrane patches of different size and the same lattice parameter also shows no significant finite-size effect.
\begin{table}[h]
	\centering
	\caption{Parameters of the least squares fitting of functions $C \left(q L\right)^{n}$ and $\left (1/\kappa\right) \left(q L\right)^{-4}$ to the thermal undulations power spectrum, ${\left\langle\tilde{h}(\mathbf{q})\,\tilde{h}^*(\mathbf{q}) \right\rangle}/{L^2}$, for membrane patches with the lattice parameter of \SI{10}{\nano\meter}, the same lateral size of $\sim$\SI{1}{\micro\metre}, and different bond-flipping frequencies, $\phi$ (data points presented in Fig. \ref{fig:th_1}).}
	\begin{tabular}{*3{>{\centering\arraybackslash}p{0.25\columnwidth}}}
		\toprule
		$\phi \, (\mathrm{ns}^{-1})$ & $n$ & $\kappa \, (kT)$\\
		\hline
			0.0 & -4.00 $\pm$ 0.04 & 17.54 $\pm$ 0.10 \\
			15.0 & -4.04 $\pm$ 0.04 & 16.79 $\pm$ 0.10 \\
			30.0 & -4.04 $\pm$ 0.04 & 17.54 $\pm$ 0.10 \\
		\hline
	\end{tabular}
	\label{tab:table_1}
\end{table}

\begin{table}[h]
	\centering
	\caption{Similar to Tab. \ref{tab:table_1} for membrane patches with the given lattice parameter $a$ and lateral dimension $L$ (data points presented in Fig. \ref{fig:th_2}).}
	\begin{tabular}{*4{>{\centering\arraybackslash}p{0.21\columnwidth}}}
		\toprule
		$a \, (\mathrm{nm})$ & $L \, (\mathrm{\mu m})$ & $n$ & $\kappa \, (kT)$\\
		\hline
		15.0 & 1 & -4.08 $\pm$ 0.13 & 19.27 $\pm$ 0.18 \\
		15.0 & 1.5 & -4.1 $\pm$ 0.09 & 19.48 $\pm$ 0.13 \\
		20.0 & 1 & -4.03 $\pm$ 0.19 & 21.86 $\pm$ 0.24 \\
		20.0 & 2 & -3.98 $\pm$ 0.11 & 22.69 $\pm$ 0.15 \\
		\hline
	\end{tabular}
	\label{tab:table_2}
\end{table}
\subsection{Area compressibility}
\label{sec:area_comp}
\begin{figure}[t]
	\includegraphics[width=\columnwidth]{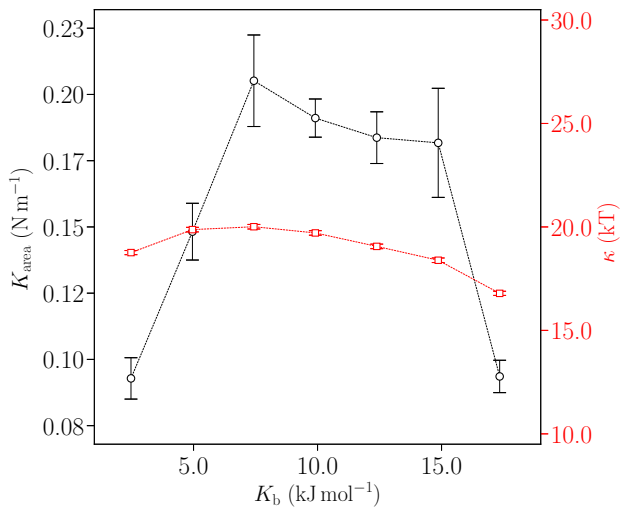}
	\caption{Isothermal area compressibility modulus and bending rigidity as functions of the angle-bending potential parameter, $K_{\mathrm{b}}$, for a membrane model with the lattice parameter of \SI{10}{\nano\meter}, equilibrated at $T=$ \SI{298}{\kelvin}. As for the bond-stretching potential parameters, if the given points are considered from left to right, respective values of $D_{\mathrm{e}}$ are \SI{173.9}, \SI{136.4}, \SI{111.2}, \SI{87.89}, \SI{64.59}, \SI{41.28}, and \SI{17.98}{\kilo \joule \per \mol}. Values of $\alpha$ are determined to be \SI{0.114}{\per \nano \meter} and \SI{0.119}{\per \nano \meter} for the first and the second points from the left, and \SI{0.120}{\per \nano \meter} for the rest.}
	\label{fig:ac_1}
\end{figure}

As an example of additional physical properties of the model that can be taken into account when choosing potential parameters, area compressibility of the membrane is calculated for a model for which potential parameters are chosen from different families. The isothermal area compressibility modulus is given in terms of the projected area fluctuations of a tensionless membrane patch as \cite{Chacon2015, Marrink2001},
\begin{equation}
\label{eqn:area_comp_fluctuation}
K_{\mathrm{area}}=A_\mathrm{eq}\left(\frac{\partial \gamma}{\partial A}\right)_T=\frac{kT \, A_\mathrm{eq}} {\langle A^2\rangle - \langle A\rangle ^ 2}
\end{equation}
where $K_{\mathrm{area}}$ is the area compressibility modulus, $\gamma$ is the surface tension, and $A_\mathrm{eq}$ and $A$ are the equilibrium and instantaneous projected areas of the membrane patch, respectively. The parameter-space optimization procedure described in Sec. \ref{sec:parameter_space_opt} is repeated for a \SI{10}{\nano\meter} model, while different bounds are put on the angle-bending potential stiffness. This yields several families of potential parameters (including the values given in the first three rows of Tab. \ref{tab:table_param}). For each set of parameters, small membrane patches have been constructed and equilibrated at $T=$ \SI{298}{\kelvin} in a tensionless state. After reaching equilibrium, fluctuations in the projected area are measured and are used to calculate area compressibility modules according to eq. \ref{eqn:area_comp_fluctuation}. Also, following the procedure described in Sec. \ref{sec:thermal_undulations}, the bending rigidity, $\kappa$, is also measured for each model. The results are ordered as functions of the stiffness of the angle-bending potential, $K_{\mathrm{b}}$, and are depicted in Fig. \ref{fig:ac_1}. It is interesting to observe that while the bending rigidity produced by various sets of parameters lie in close vicinity of the input value of \SI{20}{} $kT$, the area compressibility modulus varies considerably. Comparing the values of area compressibility modulus with the experimental range of \SIrange[range-units=single]{0.180}{0.330}{\newton \per \meter} for POPC membranes, or available simulation results (e.g. \SI{0.272}{\newton \per \meter} for POPC \cite{Janosi2010}, \SI{0.277}{\newton \per \meter} for DOPC \cite{Braun2013}, and \SIrange[range-units=single]{0.193}{0.267}{\newton \per \meter} for DPPC \cite{Klauda2010, Raghunathan2012}), shows that the model can indeed be calibrated to reproduce accurate area compressibility moduli.
\subsection{In-plane fluidity}
\label{sec:shear_flow}
\begin{figure}[t]
	\includegraphics[width=\columnwidth]{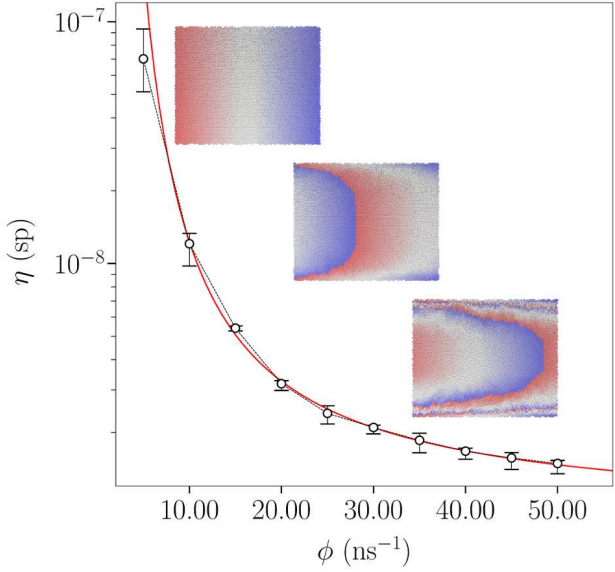}
	\caption{Surface viscosity of the membrane model as a function of the frequency of bond-flipping moves at $T$ = \SI{298}{K}. Superimposed simulation snapshots show the development of Poiseuille flow under a gravity-like force for the case of $\phi=$ \SI{5}{\per\nano\second}. The color gradient corresponds to the initial position of particles in the flow direction. The solid red line is the function $\eta = \eta_{\infty}\exp\left(C_{\phi}/\phi\right)$ fitted to the simulation results.}
	\label{fig:pf_1}
\end{figure}
As explained in Sec. \ref{sec: bondflipping}, bond-flipping Monte Carlo moves have been implemented to model the in-plane fluidity of the bilayer membrane. It is expected that the frequency of proposing bond-flipping moves, $\phi$, is correlated with the actual fluidity of the membrane. The fluidity of the 2D liquid is described in terms of the surface viscosity, which arises from the assumed linear relation between the in-plane shear stress and the corresponding velocity gradient. This assumption in effect means that the bond-flipping moves give rise to a Newtonian fluid. In order to measure the surface viscosity of the membrane, simulation of a 2D Poiseuille flow under the influence of a gravity-like force $\mathbf{f} = (mg, 0)$ with fixed parallel boundaries is performed \cite{Noguchi2005a}. The whole membrane patch is kept in a planar configuration by adding harmonic penalty for displacement in the normal direction. In order to use the Nos\'e-Hoover chains for thermostatting in this non-equilibrium setup, corresponding corrections to particle velocities are applied relative to the center of mass velocity. Reaching steady-state, the velocity component in the flow direction develops into the well-known parabolic profile with $v_{\mathrm{max}} = \rho g L^2 / 8 \eta$, where $\rho$ and $\eta$ are the membrane's surface density and surface viscosity, respectively. Superimposed frames on Fig. \ref{fig:pf_1} show development of the flow in the simulated model for the case of $\phi=$ \SI{5}{\per\nano\second} at $T$ = \SI{298}{K}. Values of surface viscosity (in units of surface poise) versus the bond-flipping frequency are also given in Fig. \ref{fig:pf_1}. As is expected, surface viscosity of the membrane decreases rapidly as the frequency of bond-flipping moves increases from \SI{5}{\per\nano\second} to \SI{50}{\per\nano\second}. The red line is a least squares fit of the function $\eta = \eta_{\infty}\exp\left(C_{\phi}/\phi\right)$ to the simulation results with $\eta_{\infty}=$ \SI{8.65e-10}{sp} and $C_{\phi}=$ \SI{26.53}{\per\nano\second}. Surface viscosity of different phospholipid bilayers have been measured to be in the range \SIrange[range-units = single]{e-7}{e-5}{sp} \cite{Camley2010, Petrov2008,Cicuta2007, Dimova1999}. If we employ this exponential fit, choosing the bond-flipping frequency in the range $\phi = $ \SIrange[range-units = single]{2.8}{5.6}{\per\nano\second} reproduces the experimental range of surface viscosities. It is to be noted that in principle, surface viscosity is a kinetic property dictated by the in-plane dynamics of the model, which in turn depends on the time integration scheme. Thus, the general procedure described in this section has to be repeated if another integrator is used.  
\subsection{Nanoparticle wrapping}
\begin{figure}[t]
	\includegraphics[width=\columnwidth]{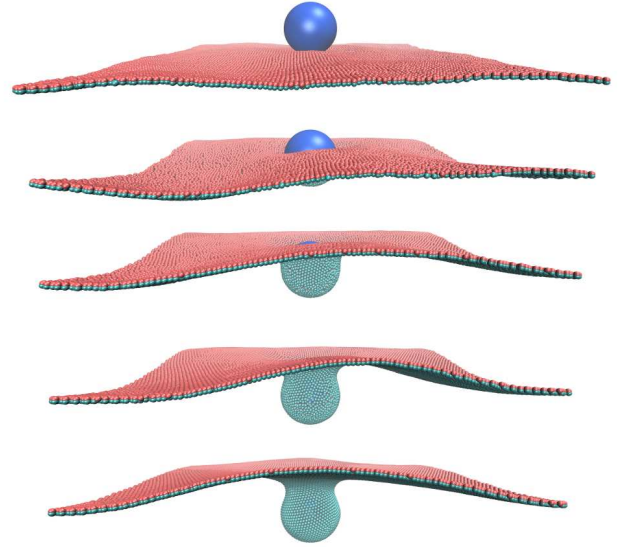}
	\caption{Representative snapshots of a nanoparticle wrapping simulation for a \SI{100}{\nano\meter} spherical nanoparticle with the dimensionless adhesion energy of $u=3.0$ and interaction range of $\rho = 0.1 R$.}
	\label{fig:np_wrapping_1}
\end{figure}

\begin{figure}[t]
	\includegraphics[width=\columnwidth]{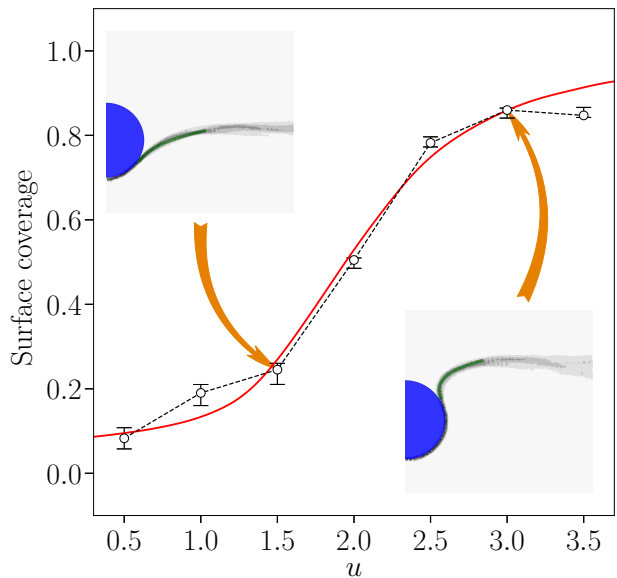}
	\caption{Fraction of nanoparticle's surface engulfed by the membrane as a function of dimensionless adhesion energy, $u$, for the same interaction range of $\rho = 0.1 R$. The continuous red line represents the prediction of the continuum model \cite{Raatz2014}. Superimposed are two slides showing ``heat maps'' of particle positions in final stages of nanoparticle wrapping. The green curves in the slides are catenary curve fits to the neck regions, corresponding to zero energy surfaces.}
	\label{fig:np_wrapping_2}
\end{figure}
As a final test of the usefulness of our membrane model to handle substantial deformations and model biologically relevant membrane remodeling processes, we simulate the interaction of spherical nanoparticles with the membrane, as a well-known benchmark system \cite{Deserno2004a,Zhang2009,Dasgupta2013, Dasgupta2014, Raatz2014}. This simple system mimics the endocytosis of nanoparticles or viral capsids by cell membranes. It is a useful test for the membrane model to show that a) the model offers enough flexibility to simulate the budding behavior of bilayer membranes, and b) if it correctly reproduces the interplay between bending and adhesion energies. For this computational experiment, a spherical nanoparticle with the radius of $R$ is put in the simulation box in the vicinity of a square shaped membrane patch. The nanoparticle interacts with the membrane through a Morse-type surface adhesion energy density of $U_{\mathrm{p}}\exp\left[-\left(r-R\right)/\rho\right]$, where $r$ is the radial distance between the center of the nanoparticle and the membrane surface. For this type of nanoparticle-membrane interaction, and with a continuum membrane model, semi-analytical studies\cite{Raatz2014} have been carried out on the degree to which the surface of the nanoparticle is covered by the membrane, as a function of the dimensionless adhesion energy $u=U_{\mathrm{p}} R^2/\kappa$ as well as the potential range, $\rho$. The parameter $u$ is the ratio between the nanoparticle-membrane adhesion energy and the energy needed to bend the membrane to a spherical shape. In order to obtain an approximation of minimum energy configurations of the membrane, and make a more meaningful comparison of the results with analytical models, simulated annealing is performed on the system of nanoparticle interacting with the membrane. The temperature of the system is decreased from \SI{300}{\kelvin} to \SI{50}{\kelvin} in 25 consecutive steps, and the system is equilibrated in each step. Fig. \ref{fig:np_wrapping_1} shows snapshots of the simulation performed for a spherical nanoparticle of \SI{100}{\nano\meter} radius interacting with a \SI{1}{\micro\meter} membrane patch, with $u=$ \SI{3.0}{}, and $\rho = 0.1 R = $ \SI{10}{\nano\meter}. Values of nanoparticle surface coverage for different choices of $u$, and for the same interaction range of $\rho = 0.1 R$ are given in Fig. \ref{fig:np_wrapping_2}. Added on the figure is the prediction of the continuum model\cite{Raatz2014} (red line). It is observed that the model follows this prediction with very good accuracy. The figure includes two slides showing ``heat maps'' of particle positions in final stages of nanoparticle wrapping for the two cases with $u=$ \SI{1.5}{} and $u=$ \SI{3.0}{}. On the slides, catenary curves are fitted to the neck region (green lines). A catenoidal membrane segment, which corresponds to zero bending energy, is expected in the unbound neck region, when the interaction range $\rho$ approaches zero \cite{Raatz2014}. Yet, for non-zero interaction range, the catenary is still a good approximation for this region \cite{Raatz2014}. The good fit to the catenary curve is an indication that the particle-based model very well captures the zero-energy regions and assumes corresponding minimal surface geometries. 
\section{Conclusion}
We have described a strongly coarse-grained model for simulating lipid bilayer membranes that is similar in nature with triangulated surface models, but is purely particle based, and as such is suitable for seamless integration into interacting-particle reaction-diffusion simulations. The model incorporates particle dimers, representing each leaflet with particles in a close-packed arrangement, and is thus suitable for distinguishing the effects corresponding to interior and exterior of cells. The lattice parameters are in the \SI{10}{\nano\meter} range, leading to each particle to laterally represent more than a hundred lipid molecules. The model relies on bond-stretching and angle-bending interactions among nearest-neighbor particles with parameters optimized to reproduce a prescribed macroscopic curvature elasticity. It is to be noted that representation of two leaflets with particle dimers also makes the inclusion of area difference elasticity possible.

It has been observed that giant plasma membrane vesicles generated from cell membranes are ``optically homogeneous'' at physiological temperatures \cite{Baumgart2007}. This in essence means that on length scales of a few hundred nanometers and above, these vesicles, which contain complex lipid and protein composition similar to cell membranes without cytoskeleta, look like and behave like reconstituted homogeneous vesicles made from a single species or a few species of lipids. Thus, whereas cell membranes have a complex composition of a multitude of lipids and proteins, which would be hard or impossible to model bottom up via molecular or coarse-grained approaches, their mechanical behavior and assumed geometries at sufficiently large length scales can be satisfactorily modeled using an elastic membrane model, such as the one proposed in this paper.

We demonstrated that the proposed model reproduces the mesoscopic physics of bilayer membranes accurately, by studying thermal undulations, area compressibility, shear flow, and nanoparticle wrapping in a quantitative manner. These computer experiments have proven the model to be reliable in different equilibrium and non-equilibrium simulations, and correctly predict the expected behavior of lipid bilayer membranes as two-dimensional fluids obeying curvature elasticity. Furthermore, it was shown that the model is tunable to include additional physical properties, such as the area compressibility modulus, via the choice of a family of potential parameters, while keeping its curvature elasticity intact. Finally, the fact that the in-plane fluidity of the membrane can be adjusted through choosing the frequency of bond-flipping moves endows the model with the ability to include regions with different viscosities, and to mimic phenomena such as lipid rafts.

The proposed model achieves remarkable computational efficiency by avoiding non-bonded pairwise interactions. With the MTK integrator used here, in the case of a \SI{1}{\micro\meter} membrane patch that constitutes about \SI{23000}{} particles, simulations done on a \SI{2.66}{\giga\hertz} machine achieved \SI{1}{\micro\second} long trajectories in less than one hour of CPU time. The small time step of \SI{20}{\pico\second} used here is due to very small particle masses. Though limiting the time step in molecular dynamics, the small masses lead to vanishing inertial contributions. Thus, if the Langevin dynamics is applied, and a stochastic integrator with significant damping corresponding to actual biological environment is used, much larger time steps are expected. This will pave the way for simulating cellular processes in their actual time scale.

Our simulations of the wrapping of spherical nanoparticles demonstrate that the presented model is able to capture biologically relevant membrane remodeling processes such as pit formation and endocytosis, where large local curvatures are induced as a result of external interactions. In contrast with the nanoparticle wrapping simulations, protein-membrane interactions can be modeled more naturally by including the induced local curvatures into the bonded interactions themselves. In PBRD or iPRD frameworks, such effects can naturally be modeled using reversible  binding-unbinding reactions. Armed with these capabilities, we ultimately aim to use this coarse-grained model in the context of iPRD simulations to study cellular signal transduction at large spatiotemporal scales.
\section{Acknowledgements}
This research has been funded by Deutsche Forschungsgemeinschaft (DFG) through grant SFB 958/Project A04 ``Spatiotemporal model of neuronal signalling and its regulation by presynaptic membrane scaffolds'', SFB 1114/Project C03 ``Multiscale modelling and simulation for spatiotemporal master equations'', and European Research Commission, ERC StG 307494 ``pcCell''.

\bibliography{library}
\bibliographystyle{aipnum4-1}

\end{document}